\documentclass[11pt]{article}

\usepackage{graphicx}

\title{Asymptotics of the number partitioning
        distribution}  

\author{Christoph Weiss and Martin Holthaus
\\
  Fachbereich Physik,\\ Carl von Ossietzky Universit\"at,
\\                D-26111 Oldenburg, Germany\vspace{1cm}
\\
pacs{05.30.Ch}: {Quantum ensemble theory}\\
pacs{05.30.Jp}: {Boson systems}  \\
pacs{02.30.Mv}: {Approximations and expansions}  
}

\newcommand{\rd}{{\rm d}}
\newcommand{\re}{{\rm e}}
\newcommand{\kB}{k_{\rm B}}

\begin{document}

\maketitle

\begin{abstract}
The number partitioning problem can be interpreted physically in terms of
a thermally isolated non-interacting Bose gas trapped in a one-dimensional
harmonic oscillator potential. We exploit this analogy to characterize,
by means of a detour to the Bose gas within the canonical ensemble, the
probability distribution for finding a specified number of summands in a
randomly chosen partition of an integer~$n$. It is shown that this
distribution approaches its asymptotics only for $n > 10^{10}$.                \end{abstract}

\begin{figure}
\centerline{\includegraphics[width=0.4\linewidth,angle=-90]{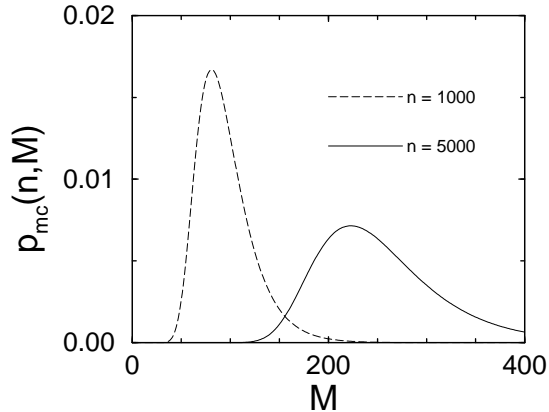}}
\caption[FIG.~1]{Exact ``microcanonical'' probability distribution~(\ref{MPD}) 
        for finding $M$ summands in a randomly chosen partition of~$n$,
        for $n = 1000$ (dashed) and $n = 5000$ (full line).} 
\label{F_1}
\end{figure}

Consider the decompositions of a natural number~$n$ into natural summands,
without regard to order. Let $\Phi(n,M)$ denote the number of such partitions 
which consist of $M$~parts, and $\Omega(n) = \sum_{M=1}^n \Phi(n,M)$ the 
total number of partitions. For $n = 4$, for instance, we have
\begin{eqnarray}
  4 & = & 1 + 1 + 1 + 1                 \nonumber \\
    & = & 2 + 1 + 1                     \nonumber \\
    & = & 2 + 2     = 3 + 1 \; ,
\label{EX4}             
\end{eqnarray}
hence $\Phi(4,4) = 1$, $\Phi(4,3) = 1$, $\Phi(4,2) = 2$, $\Phi(4,1) = 1$,
adding up to $\Omega(4) = 5$. It is known that $\Omega(n)$ grows 
exponentially with $\sqrt{n}$~\cite{HardyRamanujan18}, so that the 
enumeration of the individual partitions soon becomes impractical
when $n$ gets larger. It is then useful to focus on the distribution  
\begin{equation}
   p_{\rm mc}(n,M) \equiv \frac{\Phi(n,M)}{\Omega(n)}  
   \qquad (0 \le M \le n) \; ,
\label{MPD}  
\end{equation}
which gives the probability for finding~$M$ summands in a randomly chosen
partition of~$n$. For moderately large~$n$, this distribution can be
computed numerically with the help of the recursion relation 
\begin{equation}
  \Phi(n,M) = \sum_{k=1}^{\min\{n-M,M\}} \Phi(n-M,k) \; ; 
\label{REL}
\end{equation}     
fig.~\ref{F_1} depicts the results for $n = 1000$ and $n = 5000$.
%
%

The number partitioning 
problem~\cite{Euler11,Rademacher73,Andrews76,AhlgrenOno01} 
finds profound applications in various areas of statistical physics, 
ranging from lattice animals~\cite{WuEtAl96,BhatiaEtAl97} over combinatorial 
optimization~\cite{Mertens98} to Fermion-Boson transmutation~\cite{SM96}. 
Therefore, it is of substantial interest to characterize the 
distribution~(\ref{MPD}) for asymptotically large~$n$: Does it, 
for instance, become Gaussian?

In this Letter we tackle this number-theoretical question within a physical
framework. There is a one-to-one correspondence between the individual 
partitions of an integer and the individual microstates of a gas of ideal 
Bose particles stored in a one-dimensional harmonic oscillator potential
with frequency~$\omega_0$. If the total excitation energy~$E$ of the gas 
amounts to $n$~oscillator quanta, $E = n\hbar\omega_0$, then each partition 
of $n$ labels one possibility for distributing~$E$ among the particles. 
For $n = 4$, the first line in eq.~(\ref{EX4}) indicates a microstate with 
four quanta $\hbar\omega_0$ bestowed on four different particles, the second 
line indicates another microstate where one particle carries two quanta while 
two other particles account for the remaining ones, and so forth. For any 
value of~$E$, we assume that the number of particles be at least as large 
as the number of quanta, so that no restriction (with respect to the number 
of summands) on the partitions occurs. Hence, each excited Bose particle 
gives a nonzero summand in a partition of $E/(\hbar\omega_0) = n$, while the 
remaining ground-state particles correspond to additional zeroes.

Thus, the partitioning problem is mapped to {\em microcanonical\/} statistics: 
Given the total energy (the number to be partitioned) of the thermally 
isolated gas, the task is to count all accessible microstates. The physical 
model now suggests to consider first the simpler {\em canonical\/} version 
of this problem, {\em i.e.\/}, a gas of infinitely many, harmonically 
trapped ideal Bosons in thermal contact with a reservoir of 
temperature~$T$~\cite{Scully99}. Then the microcanonical 
distribution~(\ref{MPD}) is replaced by its canonical counterpart           
\begin{equation}
   p_{\rm cn}(b,M) \equiv \frac{\sum_n \re^{-bn} \, \Phi(n,M)}
                               {\sum_n \re^{-bn} \, \Omega(n)} \; ,
\label{CPD}
\end{equation}
where $b \equiv \hbar\omega_0/(\kB T)$ quantifies the inverse temperature, 
made dimensionless with the quantum $\hbar\omega_0$ and Boltzmann's 
constant~$\kB$. Within the canonical ensemble, the analysis starts 
from the $M$-particle partition functions
\begin{equation}
   Z_M(b) = \sum_{n=0}^\infty \omega(n,M) \, 
   \exp\!\left(-Mb/2 - bn \right) \; , 
\end{equation}
where the weight $\omega(n,M)$ is the number of possibilities for
distributing $n$~quanta over {\em up to\/} $M$~Bosons. Since $\Phi(n,M)$ 
counts the number of possibilities for distributing the $n$~quanta over 
{\em exactly\/} $M$~particles, we have
\begin{equation}
   \omega(n,M) - \omega(n,M-1) = \Phi(n,M) \; . 
\label{PHI}
\end{equation}
Following textbook practice~\cite{Pathria96}, we proceed from the canonical 
to the grand canonical ensemble by introducing the fugacity~$z$, and 
defining the grand partition function    
\begin{eqnarray}
   \Xi(b,z)  &\equiv&  
   \sum_{M=0}^{\infty} (z\re^{b/2})^M Z_M(b) 
\nonumber \\
    &=&  \sum_{M=0}^{\infty} z^M \sum_{n=0}^\infty \omega(n,M) \exp(-bn)
\nonumber \\
    &=&  \prod_{\nu=0}^{\infty} \frac{1}
         {1 - z\exp(-b\nu)} \; .
\label{GPF}   
\end{eqnarray}
If we now multiply this function~(\ref{GPF}) by $(1-z)$, so that the
ground-state factor ($\nu = 0$) is removed, eq.~(\ref{PHI}) provides
the necessary link to the desired quantities~$\Phi(n,M)$:
\begin{eqnarray}
   \Xi_{\rm ex}(b,z) & \equiv & (1 - z) \, \Xi(\beta,z) 
\nonumber \\    
   & = & \sum_{M=0}^{\infty} z^M \sum_{n=0}^\infty \Phi(n,M) \exp(-bn) 
\nonumber \\
   & = & \prod_{\nu=1}^{\infty} \frac{1}{1 - z\exp(-b\nu)} \; .
\label{GEN} 
\end{eqnarray}  
Thus, the grand partition function $\Xi_{\rm ex}(b,z)$ of an ideal Bose gas 
with amputated ground state generates the microcanonical weights $\Phi(n,M)$. 
According to probability theory~\cite{AbramowitzStegun72}, the logarithm of 
$\Xi_{\rm ex}(b,z)$ then generates the cumulants
$\kappa_{\rm cn}^{(k)}(b)$ of the canonical distribution~(\ref{CPD}):  
\begin{equation}
   \ln \Xi_{\rm ex}(b,z) = \sum_{\nu=0}^{\infty} 
   \frac{\kappa_{\rm cn}^{(\nu)}(b)}{\nu!}\left(\ln z\right)^\nu \; . 
\end{equation}
The first cumulant $\kappa_{\rm cn}^{(1)}(b)$ is the expectation value of 
the number of excited particles at the given temperature, the second
cumulant $\kappa_{\rm cn}^{(2)}(b)$ its mean-square fluctuation; in general,
$\kappa_{\rm cn}^{(k)}$ is related to the $k$-th central moment of the 
underlying probability distribution~\cite{AbramowitzStegun72}. In particular, 
$\kappa_{\rm cn}^{(k)} = 0$ for $k \ge 3$ if that distribution is Gaussian. 

It is crucial that these {\em canonical\/} cumulants can easily be
calculated in the relevant temperature regime $\kB T \gg \hbar\omega_0$, 
that is, for $b \ll 1$: Starting from the product representation~(\ref{GEN}), 
one derives~\cite{HKKS01} the {\em exact\/} formula 
\begin{equation}
   \kappa_{\rm cn}^{(k)}(b) =
   \frac{1}{2\pi i} \int_{\tau-i\infty}^{\tau+i\infty} \! \rd t \;
   b^{-t} \, \Gamma(t) \, \zeta(t) \, \zeta(t+1-k) \; ,
\label{CCF}     
\end{equation}
where $\Gamma(t)$ denotes the Gamma function, and $\zeta(t)$ is Riemann's 
Zeta function. This formula~(\ref{CCF}) allows one to determine the
asymptotic expansion of the cumulants from the residues of the integrand. 
Doing the math yields
\begin{eqnarray}
  \kappa_{\rm cn}^{(0)}(b) & \sim & \frac{\pi^2}{6b} 
        + \frac{1}{2}\ln \frac{b}{2\pi} - \frac{b}{24}
\label{KAP0}
\\
  \kappa_{\rm cn}^{(1)}(b) & \sim & \frac{1}{b}\left(\ln\frac{1}{b} 
        + \gamma \right)
        + \frac{1}{4} - \frac{b}{144} + {\cal O}(b^3)
\label{KAP1}
\\       
  \kappa_{\rm cn}^{(2)}(b) & \sim & \frac{\pi^2}{6b^2}
        - \frac{1}{2b} + \frac{1}{24}
\label{KAP2}
\\ 
  \kappa_{\rm cn}^{(3)}(b) & \sim & \frac{2\zeta(3)}{b^3} 
        - \frac{1}{12b} + \frac{b}{1440} + {\cal O}(b^3)
\label{KAP3}
\\
  \kappa_{\rm cn}^{(4)}(b) & \sim & \frac{\pi^4}{15b^4} 
        - \frac{1}{240} \; ,
\label{KAP4}
\end{eqnarray}
where $\gamma \approx 0.57722$ is Euler's constant; all higher cumulants are 
obtained in the same manner.

For applications to the partitioning problem, however, we have to abandon 
the notion of an externally imposed temperature, and to return to a thermally 
isolated gas. To this end, we write the generating function~(\ref{GEN}) as
\begin{equation}
  \Xi_{\rm ex}(b,z) = \sum_{\nu=0}^{\infty} \re^{-b\nu} \, Y(\nu,z) \; , 
\label{SER}  
\end{equation}  
where the series
\begin{equation}
  Y(\nu,z) = \sum_{M=0}^\infty z^M \Phi(\nu,M) 
\label{MAX}
\end{equation}
is of central importance, since the microcanonical weights $\Phi(\nu,M)$ 
directly figure as coefficients. Hence, its logarithm generates the cumulants 
$\kappa_{\rm mc}^{(k)}(n)$ of the microcanonical distribution~(\ref{MPD}). 
This function $Y(n,z)$ describes an ideal Bose gas which exchanges particles, 
but no energy with a reservoir, and thus coincides with the partition function 
for the recently introduced Maxwell's Demon ensemble~\cite{NavezEtAl97}. 
Writing ${\re}^{-b} \equiv x$, we extract $Y(n,z)$ from the series~(\ref{SER}) 
by means of a complex contour integral,  
\begin{equation}
   Y(n,z) = \frac{1}{2\pi i} \oint \! \rd x \, 
   \frac{\Xi_{\rm ex}(b(x),z)}{x^{n+1}} \; ,
\label{INT}
\end{equation}
where the path of integration encircles the origin of the complex $x$-plane
counter-clockwise, and evaluate this integral within the usual saddle-point
approximation~\cite{Dingle73}. The saddle point $b_0(z)$ is determined by 
setting the logarithmic derivative of the integrand to zero, resulting 
in the equation which links energy with temperature,      
\begin{equation}
  n + 1 = \left. -\frac{\partial}{\partial b} \ln \Xi_{\rm ex}(b,z)
          \right|_{b_0(z)} \; .
\label{SPE}        
\end{equation}
Within the Gaussian approximation, one is then led to 
\begin{eqnarray}
  \ln Y(n,z)  =  \ln \Xi_{\rm ex}(b_0(z),z) + n b_0(z) 
  - \frac{1}{2}\ln 2\pi
  - \frac{1}{2}\ln \! \left.\left(-\frac{\partial}{\partial b}\right)^2
  \ln \Xi_{\rm ex}(b,z) \right|_{b_0(z)} \; , 
\label{SPA}   
\end{eqnarray}
from which the desired microcanonical cumulants are obtained by further 
differentiation,
\begin{equation}
  \kappa_{\rm mc}^{(k)}(n) = \left. \left( z\frac{\rd}{\rd z}\right)^k 
  \ln Y(n,z) \right|_{z=1} \; . 
\label{DZY}    
\end{equation}
The calculations now are straightforward, but tedious, because of the 
$z$-dependence of the saddle point. We omit the technical 
details~\cite{WBHS01} and state the result for $k = 1$: When the thermally 
isolated gas carries the excitation energy $E = n\hbar\omega_0$, with 
$n \gg 1$, the expectation value of the number of excited Bose particles 
takes the form   
\begin{eqnarray}
  \kappa_{\rm mc}^{(1)}(n)  =  \kappa_{\rm cn}^{(1)}(b_1)
  - \frac{1}{2} \frac{D^2 \kappa_{\rm cn}^{(1)}(b_1)}
                     {D^2 \kappa_{\rm cn}^{(0)}(b_1)}
  + \frac{D \kappa_{\rm cn}^{(1)}(b_1)}
         {D^2 \kappa_{\rm cn}^{(0)}(b_1)}
  \left[ 1 + \frac{1}{2} \frac{D^3 \kappa_{\rm cn}^{(0)}(b_1)}
                              {D^2 \kappa_{\rm cn}^{(0)}(b_1)} \right] \; ,
\label{DIF}      
\end{eqnarray}
where $\kappa_{\rm cn}^{(0)}(b) = \ln \Xi_{\rm ex}(b,1)$, $D$ denotes the 
derivative with respect to $b$, and  $b_1 \equiv b_0(1)$ has to be taken as
function of~$n$. This latter task is achieved by inverting the saddle-point  
equation~(\ref{SPE}) for $z=1$, yielding
\begin{equation}
  \frac{1}{b_1} = \frac{\sqrt{6n}}{\pi} + \frac{3}{2\pi^2}
  + {\cal O}(n^{-1/2})  \; .
\end{equation}
Hence, the canonical expectation value~(\ref{KAP1}), expressed in terms
of the scaled energy~$n$, reads
\begin{eqnarray} 
  \kappa_{\rm cn}^{(1)}(b_1(n))  =  \frac{\sqrt{6n}}{\pi}
  \left[ \ln\!\left(\frac{\sqrt{6n}}{\pi}\right) + \gamma \right]
   +  \frac{3}{2\pi^2}\left[ \ln\!\left(\frac{\sqrt{6n}}{\pi}\right)
  + \gamma + 1 + \frac{\pi^2}{6} \right] 
   +  {\cal O}(n^{-1/2}) \; . 
\end{eqnarray}
The difference between the canonical and the microcanonical expectation
value then follows from eq.~(\ref{DIF}), utilizing the explicit expressions
(\ref{KAP0}) and (\ref{KAP1}) of the canonical cumulants
$\kappa_{\rm cn}^{(0)}$ and $\kappa_{\rm cn}^{(1)}$: 
\begin{eqnarray}
  \kappa_{\rm mc}^{(1)}(n) & - & \kappa_{\rm cn}^{(1)}(b_1(n))
   =  \frac{3}{2\pi^2}\left[ \ln\!\left(\frac{\sqrt{6n}}{\pi}\right)
  + \gamma \right] + {\cal O}(n^{-1/2}) \; .
\label{NEG}
\end{eqnarray}
Thus, we finally obtain the desired asymptotic formula for the expectation
value of the number of summands in a randomly chosen partition of a large
integer~$n$: 
\begin{eqnarray}
  \kappa_{\rm mc}^{(1)}(n) =  \frac{\sqrt{6n}}{\pi}
  \left[ \ln\!\left(\frac{\sqrt{6n}}{\pi}\right) + \gamma \right]
   +  \frac{3}{2\pi^2}\left[ 2\ln\!\left(\frac{\sqrt{6n}}{\pi}\right)
  + 2\gamma + 1 + \frac{\pi^2}{6} \right] 
  +  {\cal O}(n^{-1/2}) \; .
\label{EXP}
\end{eqnarray}
For example, for $n = 1000$ this expression~(\ref{EXP}) yields 
$\kappa_{\rm mc}^{(1)}(1000) = 94.8073...$, while the exact value is
$94.82177...$, so that the error is only about $0.015\%$.  

\begin{figure}
\includegraphics[width=0.4\linewidth,angle=-90]{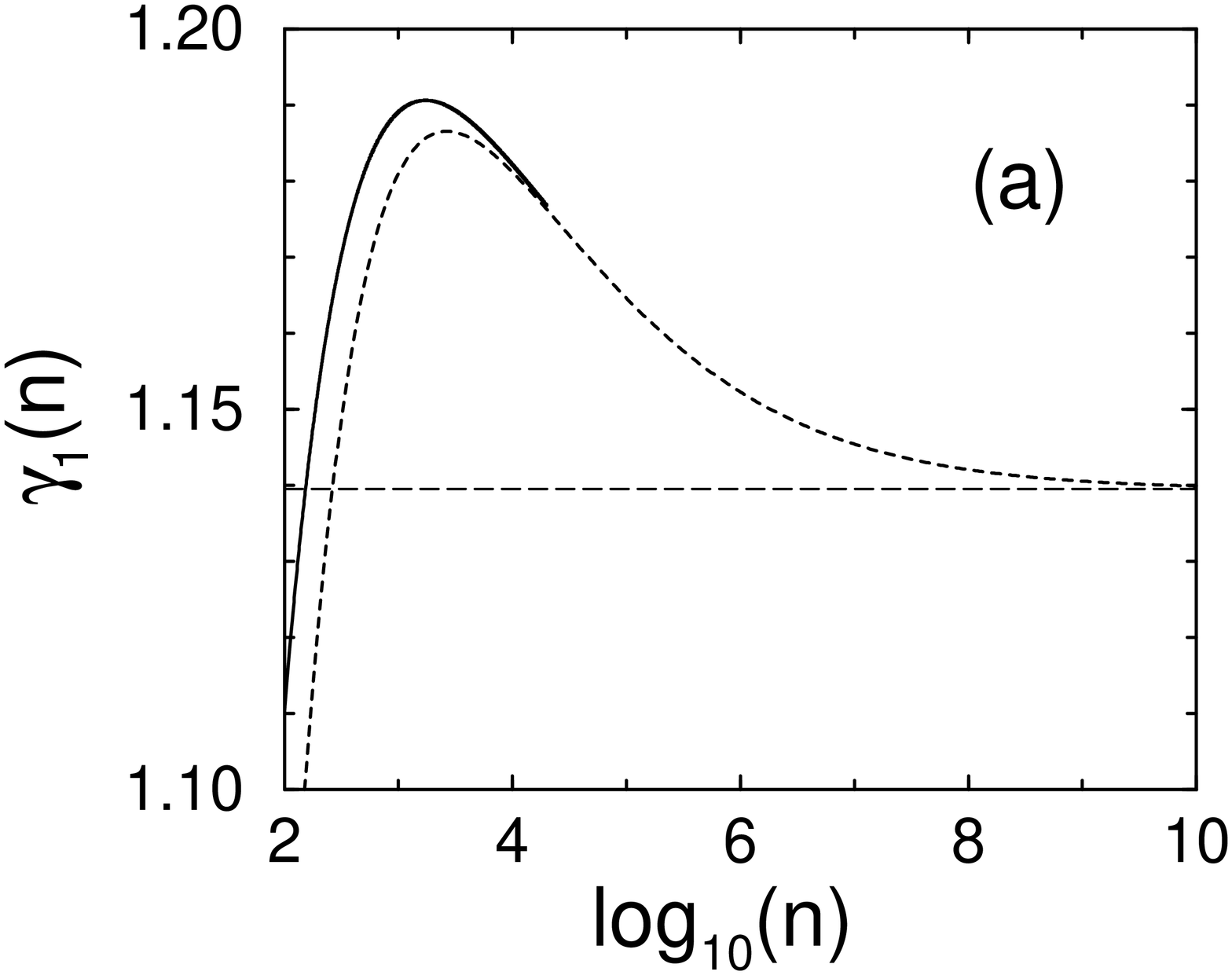}
\includegraphics[width=0.4\linewidth,angle=-90]{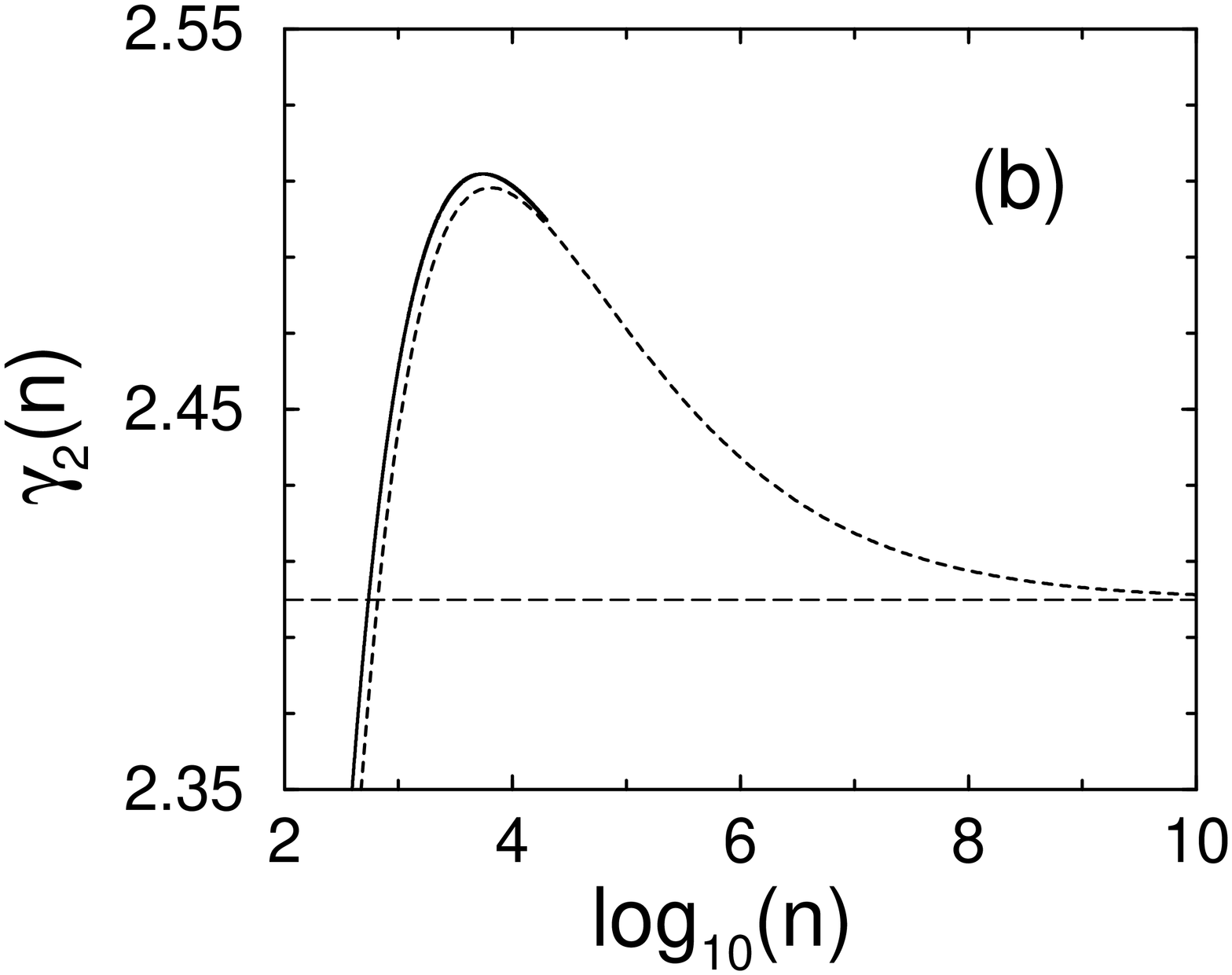}
\caption[FIG.~2]{(a)~Skewness $\gamma_1(n)$ and (b)~excess $\gamma_2(n)$
        of the number partitioning distribution~(\ref{MPD}). The solid 
        lines indicate numerically computed, exact data; the short-dashed 
        lines are the predictions of the asymptotic formulae~(\ref{SKE}) 
        and~(\ref{FLA}), respectively. The horizontal lines mark the 
        limiting values.}
\label{F_2}
\end{figure}

The above calculation of the expectation value $\kappa_{\rm mc}^{(1)}(n)$
illustrates the general strategy for computing an arbitrary 
{\em microcanonical\/} cumulant from eq.~(\ref{DZY}): Starting from the 
saddle-point approximation~(\ref{SPA}) to the generating function $\ln Y(n,z)$,
the $k$-th cumulant $\kappa_{\rm mc}^{(k)}(n)$ is expressed in terms
of derivatives of {\em canonical\/} cumulants $\kappa_{\rm cn}^{(\ell)}(b)$,
with $0 \le \ell \le k$, which, in their turn, are obtained explicitly 
from the integral fromula~(\ref{CCF}). In the same manner, the 
r.m.s.-fluctuation of the number of summands is determined as
\begin{eqnarray}
  \sigma(n)  =  \left( \kappa_{\rm mc}^{(2)}(n) \right)^{1/2}
   =  \sqrt{n} - \frac{3\sqrt{6}}{2\pi^3} \left[
  \ln\!\left(\frac{\sqrt{6n}}{\pi}\right) + \gamma + 1 \right]^2
   +  {\cal O}(n^{-1/2}) \; .
\end{eqnarray}
Of particular interest are the coefficient $\gamma_1(n)$ of skewness, and the 
coefficient $\gamma_2(n)$ of excess (or kurtosis)~\cite{AbramowitzStegun72},
\[
  \gamma_1(n) = \frac{\kappa_{\rm mc}^{(3)}(n)}
                     {\left(\kappa_{\rm mc}^{(2)}(n)\right)^{3/2}} 
  \quad \mbox{and} \quad
  \gamma_2(n) = \frac{\kappa_{\rm mc}^{(4)}(n)}
                     {\left(\kappa_{\rm mc}^{(2)}(n)\right)^{2}} \; , 
\]  
which quantify the deviation of the number partitioning 
distribution~(\ref{MPD}) from a Gaussian; in the Gaussian case,
both $\gamma_1$ and $\gamma_2$ are equal to zero. Determining the third
and fourth cumulant as outlined above, we find
\begin{eqnarray}
  \gamma_1(n) & = & 1.1395 
   +  \frac{1}{\sqrt{n}}\left[ 0.10128 \left[\ln(n)\right]^2 
        - 0.37376 \, \ln(n) - 1.7078 \right]
\nonumber \\
  & + & \frac{1}{n}\left[ 0.0075008 \left[\ln(n)\right]^4 
        + 0.025681 \left[\ln(n)\right]^3 
\right.
\nonumber \\ & &  \left.
        + 0.020024 \left[\ln(n)\right]^2 - 0.23028 \, \ln(n) - 0.56984 \right] 
   +  {\cal O}(n^{-3/2})
\label{SKE} 
\end{eqnarray}
and, again including terms of order ${\cal O}(n^{-1})$,
\begin{eqnarray}
  \gamma_2(n) & = & 2.4 
   +  \frac{1}{\sqrt{n}}\left[ 0.28440 \left[\ln(n)\right]^2 
        - 0.56714 \, \ln(n) - 10.064 \right]
\nonumber \\
  & + & \frac{1}{n}\left[ 0.025276 \left[\ln(n)\right]^4 
        + 0.022329 \left[\ln(n)\right]^3 \right.
\nonumber \\ & &  \left.
        - 0.33809 \left[\ln(n)\right]^2 + 0.73538 \, \ln(n) + 3.7863 \right] 
   +  {\cal O}(n^{-3/2}) \; .
\label{FLA} 
\end{eqnarray}
Thus, $\gamma_1(n)$ and $\gamma_2(n)$ approach nonzero constants, so that the 
distribution~(\ref{MPD}) remains non-Gaussian:  
\[
  \lim_{n \to \infty} \gamma_1(n) = \frac{12\sqrt{6}\,\zeta(3)}{\pi^3}
  \approx 1.1395
  \; ,  \quad 
  \lim_{n \to \infty} \gamma_2(n) = \frac{12}{5} \; .
\]
However, due to the vexating logarithmic corrections in eqs.~(\ref{SKE})
and~(\ref{FLA}), this asymptotic behaviour is still masked for merely
moderately large~$n$: Figure~\ref{F_2} (a) depicts exact values of 
$\gamma_1(n)$, computed numerically with the help of eq.~(\ref{REL}), 
together with the prediction of the asymptotic formula~(\ref{SKE}); 
fig.~\ref{F_2} (b) shows the same comparison for $\gamma_2(n)$. It should 
be noted that the exact evaluation of the recursion relation~(\ref{REL}) 
requires a substantial amount of computer memory and therefore becomes quite 
demanding when $n$ is of the order of $10^5$, say, while the limiting 
values of skewness and excess are well approached only for $n > 10^{10}$.  
%
%

To conclude: While the number partitioning problem is essentially 
microcanonical in nature, so that one associates ``temperature'' to
natural numbers~$n$ on the basis of their entropy $\ln \Omega(n)$, the 
equivalent problem of harmonically trapped ideal Bosons is approached 
exactly within the canonical ensemble, when temperature is imposed 
by an external heat bath. By means of such a detour to the
canonical ensemble, we have characterized the number partitioning
distribution in terms of its coefficients of skewness and excess.
Central to our approach is the fact that statistical mechanics concepts,
such as the partition function, have an intrinsic number-theoretical 
meaning~\cite{Euler11}. Even subtle differences between the two ensembles 
come into play here, such as the usually neglected difference between 
microcanonical and canonical expectation values; see eqs.~(\ref{DIF}) 
and~(\ref{NEG}). Our results show that the number partitioning distribution 
adopts its asymptotic shape only for $n > 10^{10}$, so that numerical 
simulations which inherently rely on partitions might not reach the proper 
asymptotics. The analytical method we have employed for computing 
microcanonical cumulants can be generalized to Bosons stored in different 
types of traps, and thus allows one to study the statistical mechanics of 
thermally isolated Bose gases. 

We thank M.~Block, K.~T.~Kapale, G.~Schmieder, and M.~O.~Scully for
useful discussions.

\end{document}